\documentclass[aps,preprint,showpacs,superscriptaddress,groupedaddress]{revtex4-2}  
\usepackage{graphicx}  
\usepackage{float}
\usepackage{dcolumn}   
\usepackage{bm}        
\usepackage{amssymb}   
\usepackage{slashed}   
\usepackage{amsmath}   
\usepackage{simplewick} 
\usepackage{verbatim}  
\usepackage{color} 
\usepackage{appendix} 

\usepackage[bookmarks=true,colorlinks=true,linkcolor=blue,unicode=true]{hyperref}

\begin{document}

\widetext

\title{\boldmath
Historical origins of quantum entanglement in particle physics~\footnote{ 
The Chinese version of this article was published  as   Y. Shi, {\em Historic origin of quantum entanglement in particle physics: C. S. Wu, T. D. Lee, C. N. Yang and Other Predecessors}, Micius Forum, March 17, 2023, available at   \href{https://mp.weixin.qq.com/s/gs3UxMjvXv1ert1kPu8npg}
  {https://mp.weixin.qq.com/s/gs3UxMjvXv1ert1kPu8npg}; and  also as Y. Shi, {\em Historic origin of quantum entanglement in particle physics}, Progress in Physics {\bf 43} (3), 57-67 (2023).  available at \href{https://pip.nju.edu.cn/CN/10.13725/j.cnki.pip.2023.03.001}
  {https://pip.nju.edu.cn/CN/10.13725/j.cnki.pip.2023.03.001}.  For earlier papers on contribution of C. S. Wu to quantum entanglement, see also Y. Shi,  
Chien-Shiung Wu as the experimental pioneer in quantum entanglement: a 2022 note, Mod. Phys. Lett. A {\bf 40}, 253000 (2025); Y. Shi, Scientific Spirit of Chien-Shiung Wu: From Quantum Entanglement to Parity Nonconservation,  arXiv:2504.16978} 
}
\author{Yu Shi\footnote{Yu\_shi@ustc.edu.cn}}
\affiliation{Wilczek Quantum Center, Shanghai Institute for Advanced Studies, Shanghai 201315, China}
\affiliation{University of Science and Technology of China, Hefei 230026, China} 
\affiliation{Department of Physics,   Fudan University, Shanghai 200438, China}


\begin{abstract}
In this paper, the historical origins of quantum entanglement in particle physics are systematically and thoroughly investigated. 1957, Bohm and Aharonov
noted that the Einstein-Podolsky-Rosen correlation had been experimentally  realised in  the 1949 experiment of Chien-Shiung Wu and Shaknov. This was the
first time in history that spatially separated quantum entanglement was explicitly realised in a controlled experiment. Wheeler first proposed such an experiment as a test of quantum electrodynamics, but his calculation was  in error; the correct theoretical calculations came from Ward and Price, as well as from Snyder, Pasternack
and Hornbostel, and the result was in accordance with Yang's 1949 selection rule. After the publication of Bell's inequality in 1964, it was considered whether it
could be tested by using the Wu-Shaknov experiment. This gave an impetus to the field, and a new experiment was done by Wu's group, though it was not successful as a test of Bell inequality violation. In 1957, Tsung-Dao Lee, Reinhard Oehme  and Chen Ning Yang established the quantum mechanical description of the kaons  and found that the neutral kaon is a two-state system.
In 1958, based on an approach similar to Yang's 1949 selection rule, Goldhaber, Lee  and Yang were the first to write down the entangled states of kaon pairs, in which a single kaon can be charged or  neutral. This gave, for the first time, quantum entanglement of internal degrees of freedom of  high-energy particles other than photons. In 1960, as unpublished work, Lee and Yang  discussed the consequences of quantum entanglement of neutral kaon pairs. We also describe several physicists in the past, especially Ward.
\vspace{1cm}

\end{abstract} 


\maketitle

\section{Introduction}

In 1935, A. Einstein, B. Podolsky, and N. Rosen pointed out  that local realism  
is in conflict with the completeness of quantum mechanics~\cite{1}. This is known as the Einstein-Podolsky-Rosen (EPR) paradox, and the correlation discussed is known as the EPR  correlation, which Schr\"{o}dinger coined quantum entanglement~\cite{2}. 
The original example discussed in EPR  paper was the entanglement between the positions or momenta of two particles, which are continuous variables.  
In 1951, D. Bohm gave a spin $\frac{1}{2}$  (discrete variable) version of the EPR paradox~\cite{3}.   In 1964, J. Bell suggested that local realism  leads to an inequality,  later called Bell inequality, which is violated in quantum mechanics~\cite{4}.    Later, experimental results were found to  violate  Bell inequalities, in   consistency  with quantum mechanics~\cite{4}.  

Before quantum entanglement was studied in optics, atomic physics and condensed matter physics and other areas of low-energy physics,  particle  
physics had provided concrete examples of quantum entanglement and played a certain historical role. The key players of parity revolution, especially   Chien-Shiung Wu, Chen Ning Yang and Tsung-Dao Lee, have also contributed to this less known field.  For many years, 
I have been introducing these contributions of theirs, in meetings, research papers as well as  historical  and popular articles. 
In November 2007, at the Conference in Honor of C. N. Yang's 85th birthday, I gave a presentation entitled `Professor Yang and Particle Physics", the abstract of which 
reads: ``Some of the researches  of Professor Chen Ning Yang  are 
related to quantum entangled states in particle physics.'~\cite{6}  
On the occasion of the International Symposium Commemorating the 110th Birth  Anniversary of  Chien-Shiung Wu  on 31 May 2022,  the title of my presentation was entitled `Scientific Spirit of Chien-Shiung Wu: From Quantum Entanglement to Parity Nonconservation', and the abstract reads: `In 1950, Chien-Shiung Wu and her student published a coincidence experiment on entangled photon pairs that were created in  electron-positron annihilation. This experiment precisely verified the prediction of quantum electrodynamics. Additionally, it was also the first instance of a precisely controlled quantum entangled state of spatially separated particles,  although Wu did not know about this at the time.'

More than four months after my talk on Wu's contributions,  
2022 Nobel Prize in Physics was awarded to  Alain Aspect, John Clauser and Anton Zeilinger, `for experiments with entangled photons, establishing the violation of Bell inequalities and pioneering quantum information science.'  In my review of this Prize~\cite{5},  I  mentioned two examples of quantum 
entanglement in particle physics, one being entangled photons produced by annihilation of a positron  and a electron,  and the other being entangled mesons, and  the historic role of Wu-Shaknov experiment was emphasized. Bearing the same title as this review,  I gave a talk  at the Fall Meeting of the Chinese Physical Society in November 2022, with the  
abstract including the sentence that `I would also like to take this opportunity to introduce the related work of Chien-Shiung Wu, Tsung-Dao Lee and Chen Ning Yang.'~\cite{5}

This paper provides an in-depth review of the details of this topic and its development,  clarifying some of the history and disclosing some of the lesser noticed aspects.  For example, with respect to the entangled photons produced by electron-positron annihilation, following the initial work of John Wheeler, several theoretical physicists have made important contributions to this subject.
As another example, Yang's famous 1949  photon selection rule is also closely related to the subject. The  Nobel Prize winning work on parity nonconservation  by Chen  Ning Yang and Tsung-Dao Lee   resolved $\theta-\tau$ puzzle, and identified these two kinds of particles as the same, now called kaons. Their follow-up work on kaons  laid the theoretical foundation for later discussions of meson entanglement. In 1958, M. Goldhaber, Lee and Yang discussed the kaon entangled states, giving the first entanglement of internal degrees of freedom of particles other than photons.  This is of historical significance, though they did not pay attention to the entanglement issue.  Later, in an unpublished work, Lee and Yang discussed entanglement  of kaons, referring it as EPR correlation. 
 
\section{Quantum entanglement of high energy photons}

\subsection{Entangled photons from electron-positron  annihilation}

In the 1930s, based on the Dirac equation and quantum electrodynamics, Dirac and a group of physicists studied the so-called pair theory, referring to the theory of  
 the creation and annihilation of pairs of electrons and positrons.  In 1946, Wheeler, in a paper that won an award from the New York Academy of Sciences,  systematically discussed the formation of electron-positron bound states, the simplest being a positronium.  He also discussed how to test the pair theory, suggesting that a way is to  detect the photons produced from the electron-positron  annihilation~\cite{7}.  Wheeler pointed out that the annihilation mainly comes from the spin singlet state of the positronium, i,e, the  quantum state  with total spin 0, so if the orbital angular   momentum is also 0,   the total angular momentum is  
0,   thus the linear polarisations of the two photons moving back to back from the electron-positron annihilation must be orthogonal to each other, so that the total angular momentum is  conserved. Wheeler suggested that in the experiment, each  photon is scattered separately and then detected respectively, and the events with  both photons being detected were recorded through  coincidence measurement. 

Here the photons are scattering by electrons,  known as Compton scattering. For each photon, the polarization direction determines the moving direction after scattering. So if   the polarizations of the two photons are perpendicular to each other, then  with a large probability, the moving directions are perpendicular to each other.  The  Compton scattering plays a role of polarization measurement, but as we will explain later, this `measurement'   is incomplete. 

\begin{figure}
\scalebox{0.1}{
\includegraphics{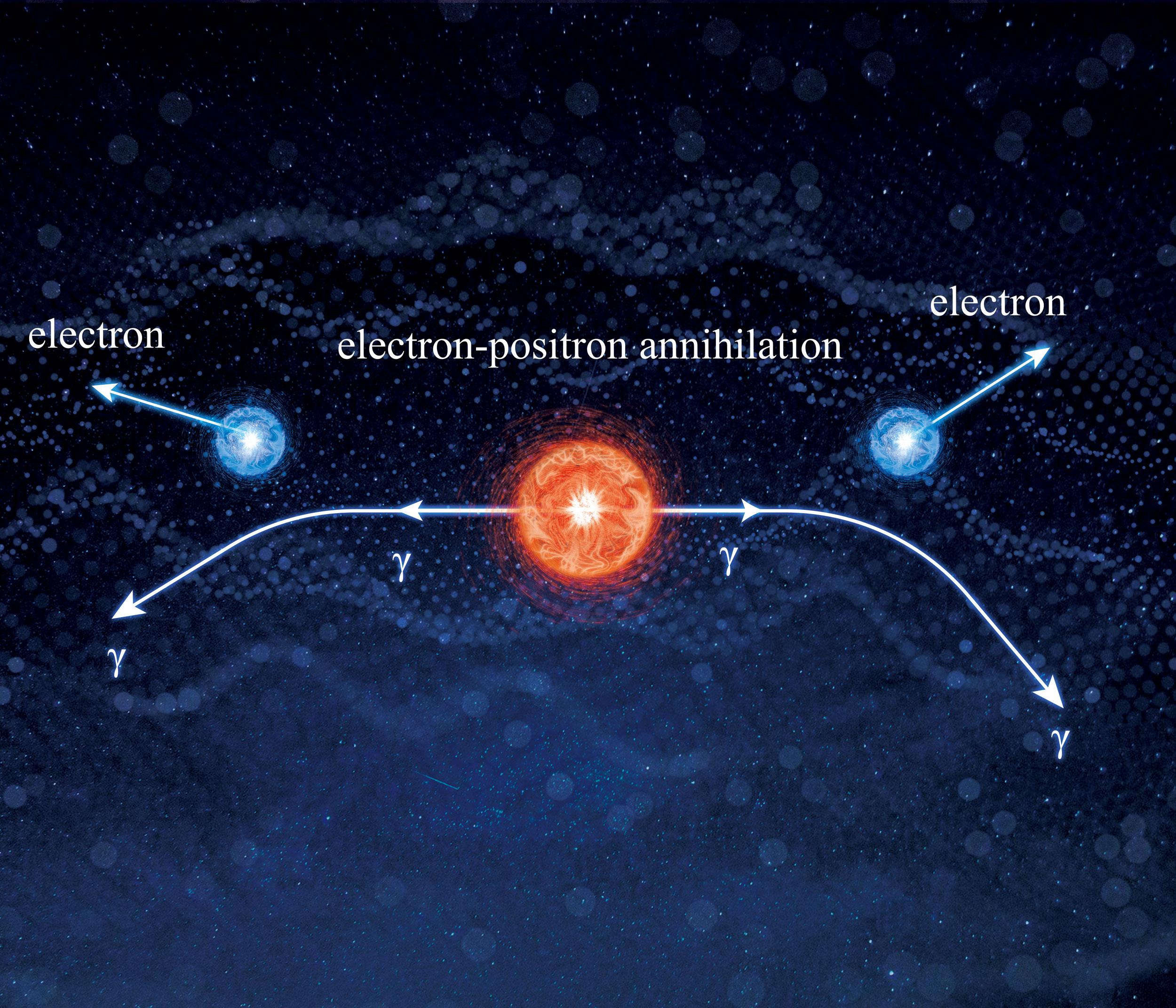}}
\caption{\label{scattering} From electron-positron annihilation, two photons are created moving to opposite directions, and are scattered by electrons in the crystals respectively.  Picture by Yu Shi at 2023.}     
\end{figure}

For each photon, the angle between the directions of motion  before and after scattering is called the scattering angle. When two photons moving in opposite directions are each scattered by an electron, even if their scattering angles are equal,   the directions of motion are not necessarily parallel, because on the plane perpendicular to the direction of motion before scattering, the azimuths can be different  (Fig.~\ref{scattering}). Wheeler suggested to study, for the case that  the scattering angles of the two photons are the same,  the asymmetry between the coincident counts of the subcase that the scattering directions are perpendicular and of the subcase that they are parallel. An asymmetry of two quantities is the difference between the two divided by the sum of the two. This asymmetry depends on the scattering angle. Wheeler calculated that when the  scattering angle is $90^{\circ}$, the asymmetry is maximal when the azimuthal difference is $74^{\circ}30'$.

Wheeler came up with the original idea, but his calculation was  erroneous. 
The correct result was given independently by two groups.
The paper by J. C. Ward and M. Pryce was received on 18 June 1947~\cite{8},  while the paper by H. Snyder, S. Pasternak and  J. Hornbostel was received on 24 November 1947~\cite{9}.  These two groups  both  calculated that   the asymmetry reaches the maximum  $2.85$  when the scattering angle is $82^{\circ}$, making correction for  Wheeler's result.  It was claimed that R. H.
Dalitz  also  obtained the result independently but did  not publish it~\cite{10}.
 
 The polarisations of the two photons produced by the electron-positron annihilation are correlated or,   in the language more commonly used today,  quantum entangled.  Wheeler did not explicitly write down the quantum state of the entangled photons, but his calculations were clearly based on the polarization entangled state, since   he made it clear that the electron and positron  are in the
spin singlet states, i.e. antisymmetric states, and that the two photons produced by the annihilation have ``similar polarization phenomena''.

But Ward and Price noted that Wheeler was mistaken about the momentum state. They published a short paper reporting only the results of the calculations, without writing explicitly  the quantum states. But
this work was part of Ward's PhD thesis~\cite{10,11,12}.   His PhD thesis
stated in details that the momentum state of the photon pair is also an antisymmetric state, which ensures that the overall state of the two photons are symmetric, obeying bosonic statistics. In today's notation, the quantum state of the  photon pair can be written as   $\frac{1}{\sqrt{2}} (|x\rangle |y\rangle - |y\rangle |x\rangle) (|\mathbf{k} \rangle |-\mathbf{k}\rangle - |-\mathbf{k}\rangle |\mathbf{k}\rangle$, where  $|x\rangle$ and $|y\rangle$  represent the two orthogonal  linearly polarised states,  and $|\mathbf{k} \rangle$ and $ |-\mathbf{k}\rangle$  represent the  
momentum states for the two opposite directions of motion respectively.

The paper by Snyder, Pasternak, and Hornbostel gave  the
correct quantum state with detailed calculations. In their abstract,  they stated  that photon scattering   acts as a ``partial analysis"  of the polarisation of the other photon. Although  a factor of 2 is missing in the number of photon pairs that are perpendicular to each other and the number of photon pairs that are parallel  to each other~\cite{11},   the antisymmetry is not affected.

\subsection{Stories of these physicists}

It is worth inserting here a little introduction to some of these physicists. 

\subsubsection{ Price}
 
Price was a student co-supervised by  M. Born and R. H. Fowler at Cambridge University in England. 
He also visited Princeton during his studies, and learnt from  
W. Pauli and J. von Neumann,   and later became Born's son-in-law. The solar neutrino conjecture  usually attributed to
B. Pontecorvo  was initially an idea of Price, when they were both working at the Chalk River Laboratories in Canada during WWII~\cite{13,14}.
In 1946, Price returned to Oxford, and Ward became his first graduate student. The problem Price suggested to Ward was to examine Wheeler's result on the electron-positron annihilation, and suggested the use of polarisation entangled states as a starting point~\cite{10,11}  Ward recalled later:``This was my first class in quantum mechanics, and actually also  the last one,   since the rest were just techniques that could be learnt from books."~\cite{10}

\subsubsection{Ward}

The other part of Ward's PhD thesis was to extend  J. Schwinger's electron self-energy renormalisation from first order to  all orders~\cite{10,11}. After a year as a tutor at the University of Sydney, he returned to Oxford to defend his  PhD thesis, followed by   two years of  research at  Oxford,  discovering the Ward's identity, which became  his most famous work, showing that renormalisation
succeeded because gauge  invariance connects different infinities. This is  a profound result and became an important element in quantum field theory.
Then he visited the Institute for Advanced Study in Princeton  for a year. He was listening to a seminar  on the two-dimensional
Ising model when he had the idea of using  combinatorial methods for this.
He published a paper on this with M. Kac in 1952.

On 6 March 2023, I asked Prof. Yang: ``in 1952, while in Princeton IAS, Ward collaborated with Kac on Ising model, by developing a combinatorial formulation. 
It is not clear how this work was influenced by your work on Ising model and phase transition. Did he ever tell you about this work?'' Prof. Yang immediately answered, ``He  did. And  I quickly wrote a paper based on his work.. "   Yang was referring to the fact that in the paper on  the  unit circle theorem of phase transitions by him and Lee,  they extended the Kac-Ward method from zero magnetic fields to imaginary magnetic field, which became an example for  the zero-point distribution of the grand  partition function  discussed in their paper.   Interestingly,  Yang's  idea also arose when listening to a seminar on   the two-dimensional Ising model, this time on the Kac-Ward method~\cite{15,16}.  I asked Prof. Yang:  ``who gave this seminar?"  Prof.  Yang replied: ``By both."   
 
Every job of Ward had not been long before he became  a professor at Macquarie University in Australia at 1967. 
In 1955, he returned to the UK to work on the hydrogen bomb project. After being given the tip ``fission, then fusion, then neutron shielding", he re-discovered   the Uram-Teller  design  of four years earlier in the United States, especially the radiation implosion. He    returned to the United States the following year~\cite{10,11}.  Ward's crucial contributions has  never been officially recognised by the UK government, although both himself and A. Salam wrote to Margaret Thatcher about it.

Around 1960, Ward (then in the USA) collaborated with Salam (then in UK) on gauge  field theory.  In 1961,  they proposed a SU(3) theory of strong interactions. And in 1964, he obtained the U(1)$\times$SU(2) electroweak theory, which  Glashow had obtained  three years earlier.  Weinberg presented the U(1)$\times$SU(2)  electroweak theory with spontaneous symmetry breaking in 1967, and Salam  proposed a similar theory  in a class in the same year, and in a Nobel Symposium in the following year.   Glashow, Salam, and Weinberg  shared the 1979 Nobel Prize in Physics~\cite{17}.  

On 27 July  2021,  I asked Prof. Chen Ning Yang about the 1979 Nobel Prize in Physics. Prof. Yang mentioned that ``Glashow's prize was based on his paper in early 1960s  proposing $SU2 \times U1$."   I said: ``Glashow did that under the bold assumption that there the gauge particles are massive. OK, the unification scheme was already correct.  Salam said that he and Ward also did this independent. But the publication was in 1964, 3 years later.   Then he claimed he also did what Weinberg did, but only in conference.''   Yang said: ``many people suspect that Salam and Weinberg got together, and decided to cut Ward out.
In the early1990s Ward  suddenly appeared in my SB office. He complained about being left out of the Nobel. 
He also complained that England did not acknowledge his contribution to the Brittish hydrogen bomb. 
At the IAS in the early 1950s I was the one  who greatly appreciated his originality."   In February 2022, I mentioned again: ``
J. C. Ward claimed that he was responsible for the design of hydrogen bomb of UK." 
Yang replied: ``He did say that, when he was quite old."  

Ward had important achievements,  although he published only about twenty papers in his lifetime~\cite{10,11,12}.

\subsubsection{Snyder}

Snyder was a student of J. Oppenheimer. In 1939, they proposed the ``continuous gravitational collapse"~\cite{18}. 
In 1947, Snyder published a paper on quantised spacetime~\cite{19}, which, incidently,  was further discussed by Chen Ning Yang   as a PhD student in the same year~\cite{20}. In the 1950s, Snyder, together with E. D. Courant and Livingston,  proposed the principle of strong focusing of sychrotrons~\cite{21,22},   which was used at CERN and Brookhaven Laboratory.  Snyder died at the age of 49~\cite{23}.

\subsubsection{Pasternak}

Pasternak was one of the first theorists to focus on the phenomenon that came to be known as the Lamb shift~\cite{24}. 
In 1934, W. Houston and Y. M.  Hsieh of the California Institute of Technology discovered that  the Balmer line series of the hydrogen atom spectrum (the spectral lines emitted when an electron jumps from a higher energy level to the second level)   deviated from the prediction of the Dirac equation. Inspired by Oppenheimer and N. Bohr, they correctly pointed out that this comes  from the self-energy of electrons due to coupling with the  electromagnetic field. R. C. Gibbs and R. C. Williams of Cornell University observed the same phenomenon and attributed the cause to the shift of the  zero angular momentum energy level   in the second shell (2s).  
In 1938, while working on his Ph.D. at Caltech, Pasternak, after discussions with Houston, also made  the same conclusion that the zero angular momentum  energy level of the second shell layer (2s) shifts, but attributed the cause to the  electron-nucleus interaction.   Later, this phenomenon was even called the Pasternak effect, which inspired W. Lamb  and R. Retherford to  use high-precision microwave techniques to measure the  difference between the energy level with zero angular momentum (2s) and that with angular momentum quantum number 1 (2p) in the second shell, later known as the Lamb shift~\cite{25}.
Lamb was awarded the Nobel Prize for this. Pasternak later became an editor of the Physical Review~\cite{26}.

\subsubsection{Wu-Shaknov experiment}

In 1949, Wu and her student I. Shaknov studied the quantum-entangled photons produced from the electron-positron annihilation   by measuring  their angular correlation after respective scattering~\cite{27}, and confirmed the theoretical  predictions of   Wheeler, Ward-Pryce, and  Snyder-Pasternak-Hornbostel.

Prior to the work of Chien-Shiung Wu and Shaknov, theoretical studies triggered  experiments by at least two groups, but the experimental results were unsatisfactory and could not give a definitive  conclusions, the problem being the efficiency of the photon detectors and the experimental conditions as written in the   Wu-Saknov paper: ``The recently developed scintillation counters have proved to be reliable and efficient gamma-ray detectors."~\cite{27} 

Wu and Shaknov have increased the efficiency of the scintillation counter, as an efficient photon detection,   to 10 times that of the Geiger-M\"{u}ller 
counter, resulting in a 100 times increase in the coincidence counting  rate. They
used two photomultiplier tubes and two anthracene crystals.  In the cyclotron in Colombia,  they bombarded  Copper 64 with deuterons to produce positrons. Then
a positron annihilated with an electron, producing two photons, which are scattered 
by electrons in the two anthracene crystals respectively. In their experiment, the mean scattering angle was very close to  $82^{\circ}$, the theoretical
value that gives the maximal asymmetry. In coincidental measurements,    one detector was kept fixed and the azimuthal angle of the other detector was taken as $0^{\circ}$, $90^{\circ}$, $180^{\circ}$ and $270^{\circ}$, respectively. The asymmetry was measured to be $2.04\pm 0.08$, which is very close to the theoretical value of $2$, calculated for their geometrical arrangement~\cite{27}. 
They gave the final words on testing the predictions of quantum electrodynamics on this problem.

\subsubsection{ Yang's selection rule}

Wu-Shaknov experiment was fully consistent with Yang's selection rule (a particle of spin 1 cannot be decay  into two photons).  In 1949, based on the invariance of rotation and inversion, Chen Ning Yang presented  the selection rule for the decay of a particle into two photons~\cite{16,28}.   
This work is also directly related to  meson decay, which we shall discuss below,  and  it also discusses the electron-positron annihilation we discussed above. 

The first sentence of this paper of Yang says that Wheeler had pointed out that a positronium in the triplet state cannot decay through annihilation with the emission of two photons.   It goes on to say that the same is true of vector and pseudovector mesons. It cites the papers that led to Wu-Shaknov experiment,  the one by Wheeler and the  two followup theoretical papers, which noted that the polarisations of the two photons are perpendicular to each other~\cite{7,8,9}.  Yang showed that these  are all consequences of the selection rules due to  invariance of rotation and that of inversion. During his time at the University of Chicago,  Yang also cited Snyder twice, once in this paper, and another time in a paper on quantized spacetime.

Yang's paper  on this selection  rule  was received on 22 August 1949 and published on 15 January 1950, while the paper  by Wu and Shaknov was received on 21 November, later than the receipt of Yang's paper, but was published on 1 January 1950, which was earlier than the publication of Yang's paper. Apparently,   they and Yang did not know each other's work at that time.

\subsubsection{Connection with the concept of quantum entanglement}

Prior to 1950, the concept of photon pairs generated by electron-positron annihilation,  including the papers by  Yang and by Wu and Shaknov,   was not connected with the concept of quantum entanglement. Now we come to the trend of quantum entanglement.  

In 1935, a few months after the publication of the EPR paper, Schr\"{o}dinger coined the term ``quantum entanglement" for  EPR correlation, but did not think it made sense. 
He believed that the EPR paradox  stemmed from taking non-relativistic quantum mechanics beyond its   range of applicability. Therefore, he also discussed the possibility that  after the separation of particles, the superposition coefficients are out of phase, and the quantum entanglement automatically disappears and the state  degenerates into  a probabilistic mixture of direct product states, i.e.,  the different direct product states appear with a certain probability. This not only avoids the EPR paradox, but also  does not contradict the experiments that had already been done at that time, which did not involve entanglement. Of course, at that time, there were no entanglement experiments, so Schr\"{o}dinger stated that this was a hypothesis. He wrote three articles on  this topic, two in English and one in German~\cite{2,29,30}.  
W. Furry also wrote two papers~\cite{31,32} examining quantum entangled states, i.e., coherent superpositions of direct product states, and probabilistic mixing of direct product states, as the two different cases. In contrast to Schr\"{o}dinger, who questioned the plausibility of entangled states, Furry  argued that it is the case of inconsistency with quantum mechanics that is implausible.
They both discuss the difference between the two cases, but only Schr\"{o}dinger's  
second English article specifically postulates that after the separation of EPR entangled pairs, their state changes  from entangled
state to probabilistic mixing~\cite{29}.  In the later literature, this assumption of  Schr\"{o}dinger and his  questioning of quantum entanglement have been frequently misunderstood as having been made by Furry. 
The timeline of the publications  of several papers by Schr\"{o}dinger and Furry is: Schr\"{o}dinger's first English paper (1935),  Schr\"{o}dinger's German paper (1935), Furry's first paper (1936),  Schr\"{o}dinger's second English paper (1936), Furry's  second paper (1936). Furry's  second paper  cites Schr\"{o}dinger's first English paper and his  German paper.  Schr\"{o}dinger's German paper  discussed measurement-induced disappearance of entanglement, and proposed the famous Schr\"{o}dinger's cat paradox~\cite{29}.

As can be seen, Einstein and Schr\"{o}dinger were worthy masters who did  not 
like the probabilistic interpretation of quantum mechanics and did not participate in the subsequent
development on this basis, but when needed, were able to make a
profound analyses within the theoretical framework of quantum mechanics. Their theoretical analyses are familiar to us today.

In 1951, Bohm gave a discrete-variable (spin-1/2) version of the EPR paradox. In 1957, Bohm and his student Y. Aharonov first connected  the EPR paradox with real physical experiments. They pointed out that in the case considered by EPR there are no interactions  between particles and their wave functions do not overlap, but at that time there was no  experimental evidence that quantum mechanics could be applied to such a many-body problem,  leading to the EPR paradox. Einstein himself, in a discussion with Bohm, said that  perhaps when the particles are separated far enough away from each other, quantum mechanics automatically fails to apply to such many-body problems~\cite{33}.

Bohm and Aharonov noted that at that time, practically discrete-variable quantum entanglement could only be studied in the polarization states of photons, which were produced in the electron-positron annihilation, and they noted that there had already been such an experiment by referring the  Wu-Shaknov paper (Shaknov was missed in the reference)~\cite{33}.   Bohm and Aharonov did not
use the term ``entanglement", but rather ``correlation". They carefully investigated the effect of correlation (entanglement) in the coincident measurement of  photon pairs after Compton scattering.  
The results showed that only entangled states can give theoretical values consistent with Wu-Shaknov experimental results, whereas the a probabilistic mixture of  direct product states discussed by Schr\"{o}dinger and Furry   leads to very different results. Bohm and
Aharonov noted only Furry's discussion and did not mention Schr\"{o}dinger's discussion.

Thus, the Wu-Shaknov experiment did produce polarization  entangled states of photons, suggesting that the EPR correlation is indeed a physical property. This was the first time in history that an  
 explicit and spatially separated quantum entanglement was achieved. In today's notation, this quantum entangled state is     
 $\frac{1}{\sqrt{2}} (|\rightarrow\rangle|\uparrow\rangle- |\uparrow\rangle|\rightarrow\rangle)$.
So the Wu-Sakhnov experiment not only accurately verified a prediction of  quantum electrodynamics,  but also became a pioneer of quantum entanglement experiments.

In 2015,  Chen Ning Yang pointed out that Wu-Shaknov experiment ``was the first experiment on quantum entanglement, which is a very hot new
area of research in the 21st century"~\cite{34}. Since most quantum states underlying physical phenomena are entangled, I would like to emphasize, as in the abstract, is that Wu-Shaknov experiment
was the first experiment explicitly realizing spatially separated quantum entanglement. 
 
\subsubsection{Can the Wu-Shaknov experiment  be used to test Bell inequality?}

Bell  inequality, published in 1964, is an inequality satisfied by several correlation functions calculated under the assumption of local realism.   In the case of photons, for example, each correlation function describes the  correlation between polarisation components of two photons in different directions. In order for the correlation functions to violate Bell inequality, the two chosen directions cannot be parallel or perpendicular, but at other angles.

Is it possible to test Bell inequality by  using the setup of Wu-Shaknov experiment? After the publication of Bell inequality, some physicists did look into this question and found that it  would not work. 

A. Shimony and M. Horne noted that in the Wu-Shaknov experimental setup, the photon polarisations detected on both sides are either parallel or perpendicular to each other,   and cannot be changed to other angles~\cite{35}.

There is another problem, the polarisation of photons in the Wu-Shaknov experiment is ``measured" through Compton scattering, but the direction of scattering is described in terms of a wavefunction, and there is a probability distribution over all directions, without  locking  to a particular direction, although the probability is maximal in the direction perpendicular to the polarisation.  Therefore the coincidence of  the photon pair does not fix the polarisation direction, and is not a perfect measurement. 

Moreover, the Wu-Shaknov experiment studied high-energy photons, the polarisations of which  cannot be measured by polarisers and polarising beam splitters  as in the case of low energy photons. Such devices  would be broken  by high-energy photons. Later on,  quantum entanglement of photon polarisation was mainly studied by using  low-energy
photons, as  in atomic physics, optics, condensed matter physics and other fields,  becoming  an important part of quantum information science and flourishing,  and three physicists received the 2022 Nobel Prize in Physics for their work in this area.

Of the three Nobel Laureates, J. Clauser received the prize in part for his work with A. Shimony, M. Horne, and R. Holt to extend Bell inequality to the CHSH inequality~\cite{36}. We note that the origin of CHSH inequality was related to their analyses of Wu-Shaknov experiment.

At that time, Clauser constructed for the Wu-Shaknov experiment a local 
hidden-variable theory~\cite{37}.  The result confirmed that  Wu-Shaknov experiment was not suitable for testing
Bell inequality. Clauser also noted the special angle between the polarizations needed for the measurements in the Wu-Shaknov experiment, and visited  Wu  to confirm this~\cite{38}.  

Clauser's visit caused Wu's interest in  testing  Bell inequality.  She and two graduate students, L. R. Kasday
and J. Ullman conducted a new experiment. This time, they
measured the coincident  probabilities of two photons at various scattering and azimuthal angles.  Their paper was completed in 1974 and published in 1975~\cite{39}.  
The paper referred to Yang's 1949 selection rule for photon pair production.

Strictly speaking, however, the new experiment of Wu's group was still not suitable for  the Bell test, because, as mentioned above, the polarisation of high-energy photons  cannot be measured perfectly, and there is always a  distribution of the scattered photons as described the wave functions. 
However, Kasday, Ullman and Wu noted that if two additional
assumptions are made that (1) the polarizations can be perfectly measured and (2) the quantum  formula for Compton scattering is correct,  then the experimental results are  consistent with quantum mechanics, and  inconsistent with Bell's inequality.

Overall speaking, the two works of Chien-Shiung Wu and her students on high-energy entangled photons, 25 years apart, contributed to  the early  study of quantum entanglement and the Bell test. 
Although they did not rigorously prove the  violation of Bell inequality,   the experimental results demonstrated quantum entanglement.  They were known to specialists on quantum foundations, for example, were referred to by John Bell in his papers. 

In 1975, M. Lamehi-Rachti and W. Mittig realized the entangled state of 
two spin-halves  originally envisioned by Bohm. They bombarded a hydrogen-containing target with a proton beam and obtained a spin singlet state consisting of two protons. Under some auxiliary assumptions, the experimental results violated the  Bell  inequality~\cite{40}.

\section{Meson entanglement}

\subsection{Lee, Oehme and Yang: neutral kaons as a quantum two-state system} 

Usually, the names of Chien-Shiung Wu,  Chen Ning  Yang and Tsung-Dao Lee are associated 
together because of parity nonconservation in  weak interactions. The 1957 Nobel Prizes awarded to  Chen Ning Yang and Tsung Dao Lee was based on their  theoretical work in 1956, which had been  initiated by the so-called $\theta-\tau$ puzzle~\cite{16,41,42}.  Parity nonconservation suggests that $\theta$ and $\tau$  are the same particle, later called kaon.  There  exist charged and neutral kaons.   They are pseudoscalar particles,   in the sense that the quantum state changes  sign under  spatial inversion.   Other similar pseudoscalar mesons include B mesons, D mesons, and so on.   

Interestingly, there are also two neutral kaons,  each one of  which is the  antiparticle of the other,  constituting a two-state system. Here, the discrete variables are flavour or strangeness.  Two  equal weight  superposed states of a particle  and an antiparticle states  are eigenstates of  C (charge conjugation) or CP (charge conjugation and parity) . Since CP is not conserved in  weak interactions, the mass-lifetime eigenstates (e.g.,  the long-lived and short-lived states of kaons) are slightly different from the CP eigenstates.  

Kaons (and other similar mesons) can be described by using the simple
Schr\"{o}dinger equation of quantum mechanics, as started by Lee, Oehme and Yang (LOY) in 1957~\cite{43}. 
In 1955, M. Gell-Mann and A. Pais proposed that the eigenstates of C or CP are formed by superposed  states of particle state and antiparticle state. 
However, at that time, they assumed that both P and C are conserved,  implying that the production of the  eigenstates of C or CP represents the production  of  the flavor eigenstates with equal probability~\cite{44}.  LOY  considered that every discrete
symmetry may be  broken, so there exists a coherent superposition of particle and antiparticle. This truly made it analogous to spin $\frac{1}{2}$. 

In May 2014, I went to CERN to attend a workshop, and my presentation was about mesonic entanglement.
On 8 May, I wrote an email to Prof. Chen Ning Yang, saying that kaon decay and neutrino oscillation can be described as simple quantum mechanical two-state or three-state systems, under Wigner-Weisskopf approximation, asking , `Are these approaches started by you?' 
 The Wigner-Weisskopf approximation is an approximation that makes the decay  exponential  with time. In 12 hours,  Prof. Yang replied, `Yes, the whole mixing matrix idea was initiated by the LOY paper,
[57e]. We used the Weisskopf-Wigner formalism to  describe the time
evolution of a system in which all 3 discrete symmetries may be
broken. At the time, this description was not really needed, since it
was believed by everybody that K1 and K2 did not mix, (because of
Gell-Mann-Pais). We developed the general case of mixing for
completeness. After1964, our formalism became THE FORMALISM. It was
generalized later to the 3 neutrino case.'

\subsection{Goldhaber, Lee and Yang: the earliest written meson entangled states}

Entangled states of mesons also first appeared in a paper by them together with Goldhaber, although they did not pay attention to the issue of quantum entanglement.  In 1958, Goldhaber, Lee and Yang  first discussed the quantum state of a pair of  K mesons $(\theta)$ \cite{45}.  
However, they considered that each particle can be in four basis states, two neutral states and a positive and negative unit charge states. 
 Although they did not discuss   from the perspective of quantum entanglement,   these two-particle
are indeed all entangled states, and four of the  entangled states are superpositions  of two-particle product states with$0$  total electric charge. .

It is worthwhile to note that the method of obtaining the entangled states of the internal degrees of freedom of these mesons is similar to the one used by Yang in giving the selection rule in 1949.  The latter is limited to 
to the production of two photons in the presence of electromagnetic or strong interactions based on conservations of  
angular momentum and parity,  while the quantum states of the meson pairs  are  based  on the conservations of strangeness,  charge conjugation and isospin for strong interactions, in the same way  as the  1949 selection rule.   The conservation of the total  variable  of the pair 
naturally leads to a variety of possibilities for individual  particles,  therefore  the total state is likely entangled.

It is interesting to note that their paper reads, `We shall show that by the combined use of
the isotopic-spin rotation operator and the charge
conjugation operator, there exist some interesting correlations,
not only in production but also between some
of the decay modes of the $\theta$ and $\bar{\theta}$'~\cite{45}.   Production  is in the basis of strangeness,   while `decay mode'  is in CP  
basis. The authors  wrote  each quantum state on both bases,  and noted correlation in  each basis.  The `interesting correlations' are  quantum entanglement.  So Goldhaber, Lee  and Yang touched on the nature of quantum entanglement.  

On 10 February 2012, I told Prof. Yang:   ``I am writing a paper on something about some analyses on entangled (EPR correlated) kaon pairs, a subject which can be traced to your paper Goldhaber-Lee-Yang 1958 on $\theta-\bar{\theta}$. Nowadays, in $\phi$ factory of Italy, kaons are produced as EPR pairs.'  Yang answered: `Please send me a copy of your paper."
 
\bibitem{Lee and Yang: Entangled States of Neutral Kaons}

In the entangled states of kaons written down by Goldhaber, Lee and  Yang,  there is superposition between charged and neutral states, with 4 of the entangled states being superpositions of states with positive and negative unit charges and neutral particle and antiparticle states.   
If we forbid the quantum coherence between charged and neutral states,  as a  superselection rule, then all these 4 entangled states reduce to  the antisymmetric superposition of neutral particle and antiparticle states, similar to spin singlet states.

According to a review paper by D. R. Inglis in January 1961~\cite{46},   in a meeting of the ZGS Users Group at Argonne National Laboratory on 28 May 1960,   Lee   discussed the possibility of correlated kaons similar to EPR question,  resulted from   proton-antiproton annihilation. 

The Inglis paper  includes a chapter on the unpublished work of Lee and Yang, which gave the 
entangled state of neutral kaons similar to the spin singlet state, where the $K^0$ and $\bar{K}^0$  are analogous to spin up and spin down, respectively, and from this one can calculate the probability that both particles are  $\bar{K}^0$.  According to this paper,  
Lee and Yang noted that it is impossible for the two neutral kaons to be observed as both $K^0$'s or both $\bar{K}^0$'s at a same time.  They also calculated the probability that the two particles are observed to be both $\bar{K}^0$'s at different moments.

The unpublished work of Lee and Yang  is referenced in this paper  as the following:  T D. Lee and C.N. Yang (unpublished); Professor Lee personal communication by letter and at a meeting of the ``ZGS Users Group" at Argonne, May 28, 1960). 

A paper  by T. B. Day, also published in January 1961, extended on the unpublished work of Lee and Yang~\cite{47}, with the citation: ``
T. D. Lee and C. N. Yang, reported by T. D. Lee at Argonne National Laboratory, May, 1960 (unpublished)."   Interestingly, Dai's article also discussed the similarity to photon pairs produced from  the electron-positron  annihilation, as we discussed above. 

My own general citation for the origin of meson entanglement is as follows: ``T.D. Lee and C.N. Yang, described in D.R. Inglis, Rev. Mod. Phys. 33
(1961) 1; T.B. Day, Phys. Rev. 121 (1961) 1204."   

In my email to Yang on 21 August 2006, I mentioned, `Recently I wrote a paper on neutral kaons (to appear in Phys. Lett. B), making a bold proposal of introducing ideas of quantum information to the realm of particle physics. I was already thinking about it when I visited you in Stony Brook three years ago.  In fact, it is ultimately based on your work with Lee circa 1960, noting that a neutral kaon pair can be created in Einstein-Podolsky-Rosen state with (J,P)=(0,-).  This work seems unpublished,  but  accounted in a paper by Inglis.'

Indeed, such neutral meson entangled states have since been widely produced and  used in meson factories~\cite{48,49,50,51,52,53,54,55,56}.  M. Jammer, in his famous book `Philosophy of Quantum Mechanics'~\cite{57}, by quoting 
Inglis and Day's article, referring to the unpublished work of Lee and Yang,  as well as Lee's report  at Argonne.

Jammer also mentioned that he had interviewed Lee  on 12 March 1973, and
he was told that Lee had noticed that  the kaon correlation is related to  EPR correlation, and was different from that of the classical ensembles. 
Jarmer wrote: ``Lee gave a talk at Argonne National
Laboratory on some striking effects of quantum mechanics in the large. In
the course of his lecture he discussed certain correlations which exist, as he
pointed out, between two simultaneously created neutral K-mesons
(kaons) moving off in opposite directions. Realizing that the situation
under discussion is intimately related to the problem raised by Einstein,
Podolsky, and Rosen, he soon convinced himself that classical ensembles
(or, for that matter, systems with hidden variables) could never reproduce
such correlations. But due to the complications caused by the finite
lifetime of kaons-for infinite lifetime the situation would "degenerate"
into that discussed by Bell-he did not derive any conclusion equivalent to
Bell's inequality but assigned the further elaboration of these ideas to his
assistant Jonas Sch\"{u}rtz, who, however, soon began to work on another
project."  Yarmer also footnoted, ``Interview with T. D. Lee, March 12, 1973. Professor Lee made it clear that all the credit should be given to Professor Bell." ~\cite{57}

In 1986, Lee published a paper entitled ``Are black holes black bodies?"   It discussed quantum entanglement,  called EPR experiment by him, across the horizon,    noting that depending on the quantum state, radiation may look like 
blackbody radiation, and  may be very different~\cite{58}. As  examples of EPR experiments and the global  nature of quantum states, the paper cites the articles by Kasday, Ullman and Wu, and by  Goldhaber, Lee  and Yang, but it did not mention  the papers by 
Inglis and by Day's, which describe the unpublished work of Lee and Yang, neither it mentioned Jammer's book or his own report at the Argonne Laboratory.

In 1996, I borrowed a copy of Jammer's book from the library of the Physics Department at Bar-Ilan University in Israel. The librarian
The librarian said, `Did you know that Professor Jammer is at our department?' 
 What a coincidence. It turns out that Jammer was   the founder of this department and was once  the president of the University. Later,  I
had some discussions with Jammer, though  didn't obtain from him any more information about the unpublished work of Lee and Yang. 

In August 2019, I emailed the article by Inglis and the relevant pages of Jammer's book to Prof. Yang. Also in this month,  Prof. Wang Chui Lin had helped me to look for written material about  Lee-Yang   unpublished work on neutral kaon  entanglement, Lee's report at the Argonne Laboratory, and first-hand accounts of his correspondence with Inglis, but none was found.

\subsection{ Friedberg's work}

According to Jammer, R. Friedberg did some unpublished work in this area~\cite{57}. Friedberg had been  a student of Lee, and remained in Columbia as Lee's   long-time collaborator.  
It is not known whether this work of Friedberg was  advised by Lee. 

In 1967, unaware of Bell's work, Friedberg
applied the assumption of locality to spin measurements, obtaining results that contradicted quantum mechanics. 
In 1968, he told Jammer about this work, and in 1969, he wrote about it in an unpublished paper:  R. Friedberg, ``Verifiable consequences of the Einstein -Podolsky-Rosen criterion for reality" (unpublished, 1969). ~\cite{57}.

Friedberg first reformulated the criterion for reality as follows. For
two systems, it is possible to measure the first system without disturbing the second system, and it is also possible to measure the second system without disturbing the
first system. If the results of the two measurements match exactly, then this
result is part of the reality, even without actually measuring it.

He then consideres that each system has three quantities $x$, $y$, and $z$, all taking the value of $1$ or $-1$.
For each system, any two quantities can be measured simultaneously, since one can  be measured directly and the other can be measured on  the EPR entangled state by measuring the other. One can thus obtain the average values of the products  satisfy $\langle xy\rangle+\langle yz\rangle +  \langle xz \rangle  \geq  -1$.  However, for quantum mechanical spin, if $x$, $y$, and $z$ correspond to the 3 components of the spin, it can be shown that they satisfy $(\langle xy\rangle  + \langle yz\rangle  +  \langle xz \rangle)^2 \geq  1$, which can violate   the inequality $\langle xy\rangle  + \langle yz\rangle  +  \langle xz \rangle \geq  -1$.   W. B\"{u}cher from  Germany obtained a similar result in 1967~\cite{57}.

Friedberg  did another unpublished work in 1969,  giving a simplified proof of the Kochen-Specker theorem~\cite{57}. The Kochen-Specker theorem states that, under non-contextual  assumption,   it is not possible to self-consistently assign a deterministic value to each observable.

For the Wu-Shaknov experiment, Friedberg has told Jammer that the unentangled  case considered by Furry could be represented in terms  of Bell  inequality, and that the corresponding correlation function is different from that for the entangled state, which is a cosine function,  which is multiplied with a  coefficient not exceeding 1/2 in giving the former~\cite{57}.

\section{Finding the 0 to 1 Trail}

We now see that the early work on quantum entanglement in particle physics
played a crucial historical role in promoting the study of quantum entanglement. Take the work of Bell for an important example~\cite{59},  his two famous earliest papers ``On the problem of hidden variables in quantum mechanics" and ``On the Einstein-Podolsky-Rosen paradox",  published in 1966 and  1964, respectively (the former had been written earlier)  cited Bohm-Aharonov  1957 papers; a 1971 paper  ``Introduction to the Hidden-Variable Question"  cited the papers by Day and  by Inglis;  a 1975 paper ``On wave Packet Rudection in the Coleman-Hepp Model"  cited Jammer's  book,  saying ``See  in particular references to T. D. Lee (p. 308) and R. Friedberg (pp. 244, 309, 324)", and cited  a contribution by Kasday to a conference together with the paper  by Kasday, Ullman and Wu. 
 
On 10 December 2022, in an email to Prof. Yang, I
said, ``Bell inequalities finally got Nobel Prize, though not to Bell himself. 
I remember you mentioned in SPI (Yang's Selected Papers I) that when you visited CERN,  you told John Bell your work on ODLRO, and Bell proved some of your conjectures.  Any more memories of this man?"   Prof. Yang  immediately replied, ``He was very good.''

On 11 March 2023, after the present paper was almost complete, I expressed my opinion to Prof. Yang, `I would like to make a point that 1958 Goldhaber-Lee-Yang paper is very important in the perspective of entanglement.  It reads: ``We shall show that by the combined use of the isotopic-spin rotation operator and the charge conjugation operator there exist some interesting correlations, not only in production but also between some of the decay modes.''  My two cents: 1. It used the same method as Yang's 1950 selection rule, and derived  kaon entangled states as the production, just as in the case of photon pair.  
2. Just as the photon pair in Yang's selection rule can be entangled, kaon pairs are entangled, and it was already noted that the decay mode are entangled (though this term was not used).    
3. Here it was not constrained that each kaon in the product must be neutral. Later people only considered neutral kaons.   This paper was the first noting that two kaons (or any kinds of particles besides photons) can be entangled.'
  
In the unpublished work by Lee and Yang   in 1960, the joint probability calculated for the neutral kaon entangled state  (and a later focus of attention of the calculation and measurement of such entangled state, analogous to the photon coincident  probability) was a manifestation of the  correlation of decay modes referred mentioned in 1958. 

We  now  emphasize the breakthrough from 0 to 1, but in history, it has not always been a quick fix.  Over time, the contributions of some scientists in the 0 to 1 process may have been forgotten, especially if those scientists are not well known. Even famous scientists may not always be remembered for their original efforts in certain areas,   especially if those fields were not well known at the time.   It is worthwhile to sort out, examine, and learn from the efforts, both successful and not-so-successful, that have been made along the way in science.

\acknowledgments

I would like to thank Prof.  Chen Ning Yang for his communications. This paper has been  supported by the National Natural Science Foundation of China (Grant No. T2241005).

\end{document}